\documentclass{czjphys}         
\begin{document}
\parskip 5pt
\newcommand{\g}{{\bf g}}
\newcommand{\h}{{\bf h}}

\def\nn{\nonumber}
\def\d{\partial}
\def\v{\vert}
\def\l{\langle}
\def\r{\rangle}

\title{Exact solvability and quantum integrability of a derivative nonlinear 
Schr${\rm {\ddot o}}$dinger model}
%
\authori {B. Basu-Mallick and Tanaya Bhattacharyya} 
\addressi {Theory Group, 
Saha Institute of Nuclear Physics, 
1/AF Bidhan Nagar, Kolkata 700 064, India}
\authorii{}     \addressii{}
\authoriii{}    \addressiii{}
\authoriv{}     \addressiv{}
\authorv{}      \addressv{}
\authorvi{}     \addressvi{}
%
\headauthor{B. Basu-Mallick and T. Bhattacharyya} 
\headtitle{Exact solvability and quantum integrability 
...}             
\lastevenhead{B. Basu-Mallick and T. Bhattacharyya
 } 

\refnum{A}
\daterec{XXX}    
\issuenumber{0}  \year{2003}
\setcounter{page}{1}
\maketitle

\begin{abstract}
By using a variant of quantum inverse scattering method (QISM) which is 
directly applicable to field theoretical systems,
we derive all possible commutation relations among the operator valued 
elements of the monodromy matrix associated with an 
integrable derivative nonlinear
Schr${\rm {\ddot o}}$dinger (DNLS) model. From these    
commutation relations we obtain the exact Bethe eigenstates for the 
quantum conserved quantities of DNLS model.
We also explicitly construct the 
first few quantum conserved quantities including 
the Hamiltonian in terms of the basic field operators of this model.
It turns out that this quantum Hamiltonian has a
new kind of coupling constant which is quite different from the
classical one. This fact allows us to
 apply QISM to generate the spectrum of quantum DNLS Hamiltonian
for the full range of its coupling constant.

\end{abstract}

\section{Introduction}

Derivative nonlinear
Schr${\rm {\ddot o}}$dinger (DNLS) model in $1+1$-dimension 
is a well-known integrable system which has found 
applications in different areas of physics like circularly polarized nonlinear 
Alfven waves in a plasma [1], quantum properties of solitons in optical
fibres [2] and some chiral Luttinger liquids [3]. 
The equation of motion for the Chen-Lee-Liu type
 classical DNLS model is given by  
\be
i \d_t \psi (x,t) + \d_{xx}\psi(x,t) - 4i \, \xi \,
{\psi^* (x,t)}{\psi (x,t)} 
\d_x{\psi(x,t)} \, = \,  0 \, , 
\label {a1}
\ee 
where $\d_t \equiv {\d \over \d t}$,
$\d_x \equiv {\d \over \d x}$,
$\d_{xx} \equiv \frac{\d^2}{\d x^2}$ and
$\xi$ is a real parameter representing the
strength of the nonlinear interaction term. The Hamiltonian 
\be
 H = \int_{-\infty}^{+\infty}
\left\{ -  \psi^*(x) \, \d_{xx} \psi(x) \, + \, i\xi \,
 {\psi^*}^2(x) \, \d_x  \psi^2(x)  \right \}
 dx \,  ,
\label{a2}
\ee
and an equal time Poisson bracket structure given by
$
\{ \psi(x), \psi(y) \} = \{ \psi^*(x), \psi^*(y) \} 
\hfil \break
 = 0,  $ 
$  \{ \psi(x),
 \psi^*(y) \} =
-i \delta(x-y) \, ,
$
 generate eqn.(\ref {a1}) as a canonical evolution. 
By using the Lax operator associated with this DNLS model,
one can explicitly construct its conserved quantities
in a recursive way [4].
It can be shown that such
 infinite number of conserved quantities 
 yield vanishing Poisson bracket relations 
among themselves. 
This fact establishes the classical integrability
of DNLS model (\ref {a1}) in the Liouville sense. 

In this review article, our aim is to study
 the spectrum for the quantum version
of the above mentioned DNLS model [5]
 and discuss the nature of the related conserved quantities [6]. In Sec.2,
we construct the Bethe eigenstates for these conserved 
quantities by using a variant of quantum inverse scattering method (QISM) [7]
which is directly applicable to continuum field models.
In Sec.3, we derive the expressions of 
the first few quantum conserved quantities including the
Hamiltonian in terms of the basic field operators
of the system. Sec.4 is the concluding section.

\section{Application of QISM to quantum DNLS model}

In the quantized version of
DNLS model, the basic field operators 
satisfy equal time commutation relations given by
$$
\Big[\psi( x ), \psi( y )\Big] = 
\left[ \psi^\dagger( x ) , \psi^\dagger( y ) \right] = 0 , ~~
\left[ \psi( x ) , \psi^\dagger( y ) \right] = \hbar\delta( x - y ) \, , 
$$
$\hbar$ being the Planck's constant, and the 
 vacuum state is defined through the relation: $\psi(x) \v 0 \r = 0$.
For applying QISM [7] to this model, it is necessary to find out
at first an appropriate Lax operator which would satisfy
quantum Yang-Baxter equation (QYBE) in continuum. Such a 
quantum Lax operator is obtained as [5]
\be
       {\cal U}_q ( x, \lambda ) =
i \pmatrix { f \, \psi^\dagger(x)\psi( x )
 - {\lambda^2}/4  &
\xi\lambda\psi^{\dagger}( x ) \cr
\lambda\psi( x ) &  - g \, \psi^\dagger(x)\psi(x)
 + \lambda^2/4 } \, , 
\label {b1}
\ee 
where $f = \frac{\xi
e^{-i\alpha/2}}{\cos \alpha / 2}$ , $g = \frac{\xi
e^{i\alpha/2}}{\cos \alpha / 2}$ and $\alpha $ is a
real parameter which is fixed through the relation:
 $\sin\alpha = -\hbar\xi$. Due to this relation, 
 it is evident that QISM is applicable to
 quantum DNLS model when the parameter $\xi $ satisfies a restriction
given by $\v \xi \v \leq {1\over \hbar}$.

Using certain symmetry properties of the quantum Lax operator (\ref {b1}),
the corresponding quantum monodromy
matrix ${\cal T}(\lambda)$ can be written in the form
\be
{\cal T}(\lambda)=\pmatrix {A(\lambda) & -\xi B^\dagger(\lambda) \cr
                          B(\lambda) & A^\dagger(\lambda)} \, .
\label {b2}
\ee
  Expressing QYBE satisfied by the above monodromy matrix
 in elementwise form, we obtain [5]
\bea
&&\Big[ A( \lambda ), A( \mu ) \Big] = 0 , ~
\Big[ B( \lambda ), B( \mu ) \Big] = 0 , ~
A(\lambda) B^\dagger(\mu) = \frac{\mu^2 q - \lambda^2 q^{-1}}{\mu^2 -
\lambda^2 - i \epsilon} B^\dagger( \mu ) A( \lambda )  , 
 \nn \\
&&B( \mu ) B^\dagger( \lambda ) = \tau(\lambda,\mu)   
 B^\dagger( \lambda ) B( \mu ) 
+ 4 \pi  \hbar  \lambda \mu \, 
\delta( \lambda^2 - \mu^2 ) A^\dagger( \lambda ) A( \lambda )  , \nn  \,   
~~(5a,b,c,d)
\eea
\addtocounter{equation}{1}
where $\tau( \lambda, \mu ) = \left[ \, 1 + \frac{8  \hbar^2 \xi^2 \lambda^2
\mu^2}{{( \lambda^2 - \mu^2 )}^2} - \frac{4 \hbar^2 \xi^2 \lambda^2
\mu^2}{( \lambda^2 - \mu^2 - i \epsilon ) ( \lambda^2 - \mu^2 + i \epsilon
)} \, \right]$.
Due to eqn.(5a), it follows that all operator
valued coefficients occurring in the expansion of
 $\ln A(\lambda)$ in powers of $\lambda $
must commute among themselves. Consequently,
 $\ln A(\lambda)$ may be treated as the generator of
conserved quantities for the quantum integrable DNLS model.
By using algebraic Bethe ansatz,
 one can also find out the eigenfunctions and
eigenvalues corresponding to different expansion coefficients of
 $ \ln A(\lambda) $. However, such eigenvalues
 would be complex quantities in general. 
To make the eigenvalues real, we define another operator    
${\hat A}(\lambda)$ through the relation: 
${\hat A}(\lambda) \equiv 
A(\lambda  e^{ -{i \alpha\over 2}}) $ and expand $\ln {\hat A}(\lambda)$ as 
\be
 \ln {\hat A}(\lambda) =
\sum_{n=0}^{\infty} \frac{ i \,{\cal C}_n}{\lambda^{2n}}  \, .
\label {b3}
\ee
With the help of eqns.(5c) and (\ref {b3}), we easily find that
\bea
&& {\cal C}_0 \v \mu_1 ,\mu_2  , \cdots , \mu_N  \r ~=~\alpha N
\,
  \v \mu_1 ,\mu_2  , \cdots , \mu_N  \r \, , \nn \\
&& {\cal C}_n \v \mu_1 ,\mu_2  , \cdots , \mu_N  \r =
\frac {2}{n} \sin(\alpha n) \Big \{ \sum_{j=1}^N \mu_j^{2n} \Big \}  \,
  \v \mu_1 ,\mu_2  , \cdots , \mu_N  \r \, , 
\label {b4}
\eea
where $n \geq 1$ and 
  $\v \mu_1 ,\mu_2  , \cdots , \mu_N  \r \equiv
 B^\dagger(\mu_1) B^\dagger(\mu_2) \cdots B^\dagger(\mu_N) \v 0 \r $ 
 represent the Bethe eigenstates for all ${\cal C}_n$s.
For the case of scattering states,  
$\mu_j$s are chosen as all distinct real or pure imaginary numbers.
We can also construct the quantum soliton states or bound states
 for DNLS model by choosing complex values of $\mu_j$ in an 
appropriate way [4,5]. For the case of both scattering and solitonic
bound states, eqn.(\ref {b4}) yields real eigenvalues for all ${\cal C}_n$s.

Thus, by applying algebraic Bethe ansatz,
 we obtain the spectra of all ${\cal C}_n$s
which are formally defined through the expansion (\ref {b3}). Next, we turn 
our attention to the explicit construction of these quantum conserved
quantities in terms of 
basic field operators like $\psi (x)$ and $\psi^\dagger(x)$.
In analogy with the cases of 
nonlinear Schr${\ddot {\rm o}}$dinger (NLS) model and sine-Gordon model, 
one may think that the Hamiltonian of quantum DNLS model
can also be obtained as the normal
ordered version of the corresponding classical Hamiltonian (\ref {a2}). 
However, since QYBE restricts the value of $\xi$  
 as: $\v \xi \v \leq {1\over \hbar}$, such a quantum 
Hamiltonian would be solvable by QISM only within 
this limited range of coupling constant. 
So at this point, our target is to find out
 explicit expressions for the first few quantum
conserved quantities of DNLS model 
and investigate whether the above mentioned limitation 
about the applicability of QISM can be resolved in some way. 

\section{Quantum conserved quantities of DNLS model}

For obtaining the quantum conserved quantities of DNLS model
in terms of the basic field operators $\psi(x)$ and $\psi^\dagger(x)$, 
here we wish to follow the approach of Ref.8 
where first few conserved quantities 
of the quantum NLS model has been constructed explicitly 
 by using the so called `fundamental relation'. 
To this end, we consider Jost solutions associated with the 
quantum Lax operator (\ref {b1}). It can be shown that 
 the components of the Jost solution 
$\rho( x, \lambda ) \equiv \pmatrix
{\rho_1( x, \lambda ) \cr
\rho_2( x, \lambda )} $ , associated with 
 boundary conditions at $x\rightarrow -\infty $, follow the
differential equations given by [6]
\bea
&&\partial_x\rho_1( x, \lambda ) \, = \,    
- \frac{i\lambda^2}{4}\rho_1( x, \lambda ) +
 if\psi^\dagger( x ) \rho_1( x, \lambda)\psi( x )
+ i\xi\lambda \psi^\dagger( x )\rho_2( x, \lambda ) \, , \nn \\
&&\partial_x\rho_2( x, \lambda ) \, = \, 
 \frac{i\lambda^2}{4}\rho_2( x, \lambda ) 
-ig\psi^\dagger( x ) \rho_2( x, \lambda)\psi( x ) 
+ i\lambda\rho_1( x,\lambda )\psi(x) \, .
\label{c1}
\eea
There exist two independent $\rho(x,\lambda)$ satisfying the above relations 
with boundary conditions like 
$\pmatrix {\rho_1( x, \lambda ) \cr
           \rho_2 ( x, \lambda )} 
 \stackrel{x \rightarrow \, - \infty}{\longrightarrow} 
\pmatrix { e^{-i\lambda^2 x/4} \cr 0}$  or 
$\pmatrix {\rho_1( x, \lambda ) \cr
           \rho_2 ( x, \lambda )} 
 \stackrel{x \rightarrow \, - \infty}{\longrightarrow} 
\pmatrix {0 \cr  e^{i\lambda^2 x/4}} \, . $ 
On the other hand, the 
 components of Jost solution 
$\tau (x,\lambda) \equiv \pmatrix {\tau_1( x, \lambda
) \cr
\tau_2( x, \lambda )} $ associated with the boundary conditions at 
$x\rightarrow +\infty$ follow the differential equations given by
\bea
&&\partial_x\tau_1( x, \lambda ) \, = \,
- \frac{i\lambda^2}{4}\tau_1( x, \lambda ) + ig \, \psi^\dagger( x )
\tau_1( x, \lambda) \psi(x) 
+ i\xi\lambda \,
\psi^\dagger(x)\tau_2( x, \lambda ) \, , \nn \\
&&\partial_x\tau_2( x, \lambda ) \, = \,
 \frac{i\lambda^2}{4}\tau_2( x, \lambda ) - if \, \psi^\dagger( x )
\tau_2( x, \lambda) \psi(x) + i\lambda \,
\tau_1( x, \lambda ) \psi(x) \, . 
\label{c2}
\eea
Again, there exist two independent $\tau(x, \lambda)$
satisfying the above relations and obeying 
  boundary conditions like 
$ \pmatrix {\tau_1( x, \lambda ) \cr
           \tau_2 ( x, \lambda )} 
 \stackrel{x \rightarrow \, + \infty}{\longrightarrow} 
\pmatrix { 0 \cr e^{i\lambda^2 x/4} } $   or
$ \pmatrix {\tau_1( x, \lambda ) \cr
           \tau_2 ( x, \lambda )} 
 \stackrel{x \rightarrow \, + \infty}{\longrightarrow} 
\pmatrix {- e^{ - i\lambda^2 x/4} \cr 0} \, . $ 
It may be noted that the sets of equations (\ref {c1}) and (\ref {c2}),
satisfied by the quantum Jost solutions defined through 
 boundary conditions at $x\rightarrow
+\infty $ and $x\rightarrow - \infty$ respectively, 
are not identical in form.
This happens due to the fact that, unlike the cases of most other
integrable systems, the quantum Lax operator (\ref {b1}) of
DNLS model is not a traceless matrix [6].
   
Next, let us define the quantum Wronskian
associated with the Jost solutions $\tau (x,\lambda)$ and
 $\rho (x,\lambda)$ as 
\be
\Lambda_{\rho , \tau }(x,\lambda) =
\tau_2( x, \lambda) \rho_1( x,\lambda)
- \tau_1(x,\lambda) \rho_2(x,\lambda) \, .
\label{c3}
\ee
All elements of the quantum 
monodromy matrix ${\cal T}(\lambda)$ (\ref {b2})
 can be expressed through this 
quantum Wronskian by taking different choices of $\rho(x,\lambda)$ 
and $\tau(x,\lambda)$ [6]. Thus the quantum Wronskian (\ref {c3})
does not really depend on the value of the coordinate $x$. 
In analogy with the quantum Wronskian,
we define another operator associated with the Jost solutions
of DNLS model as 
\be
\Gamma_{\rho,\tau}( x,\lambda )  \, =  \,
\tau_2( x, \lambda ) \rho_1( x, \lambda )
+ \tau_1( x, \lambda ) \rho_2( x, \lambda )
  \, .
\label{c4}
\ee
Next, we propose that,
the quantum conserved quantities ($I_n$) of DNLS model
would annihilate the vacuum state and obey the 
`fundamental relation' of the form 
\be
\Big[ I_n , \Lambda_{\rho,\tau}( \lambda ) \Big]
= \frac{\hbar\lambda^{2n}}{2^{n+1}} \, 
\Big \{ \, \Gamma_{\rho,\tau}( +\infty, \lambda )
- \Gamma_{\rho,\tau}(-\infty, \lambda ) \, \Big \} \, , 
\label{c5}
\ee
where $n$ is any nonnegative integer. Let us now explore 
some consequences of this fundamental relation. 
Using the fundamental relation (\ref {c5}) for different choices
 of $\rho(x,\lambda)$ and $\tau(x,\lambda)$,
we obtain the commutation relations given by [6]
\bea
~~~~~~~~~~~~~~~~~~~\Big[ I_n, A( \lambda ) \Big] = 0  \, , ~~~~
\left[I_n , B^\dagger(\lambda) \right] 
= \frac{\hbar\lambda^{2n}}{2^n} B^\dagger 
( \lambda ). \nn
~~~~~~~~~~~~~~~~~~~~(13a,b)
\eea                 
\addtocounter{equation}{1}
By using the commutation relation (13b),
 it is easy to show that $I_n$s 
 satisfy eigenvalue equations like
\be
I_n \, \v \mu_1,\mu_2,\cdots ,\mu_N \r \, = \,
 \Big( \, \frac{\hbar}{2^n}\sum_{j=1}^N \mu_j^{2n} \, \Big) \v
 \,  \mu_1,\mu_2,\cdots ,\mu_N \r \, ,
\label{c7}
\ee
where $\v \mu_1,\mu_2,\cdots ,\mu_N \r \equiv B^{\dagger}(\mu_1)
B^{\dagger}(\mu_2)\cdots B^{\dagger}(\mu_N)\v 0 \r $.
Consequently,
the commutation relation (13b) may be treated as the spectrum
generating algebra for the quantum conserved quantities of DNLS model.
Now we assume that the Bethe states
 $\v \mu_1,\mu_2,\cdots ,\mu_N \r $ represent a complete set of
states in the corresponding Hilbert space. Thus two operators  
would coincide if they can be simultaneously 
diagonalised through these complete set of states and
their eigenvalues always match with each other.
Comparing (\ref {c7}) with (\ref {b4}), it is easy to find that
\be
{\cal C}_0 \, = \, \frac{\alpha}{\hbar} \, I_0 \, ,~~~
{\cal C}_n \, = \,  \frac{2^{n+1}}{n\hbar} \sin(\alpha n) \, I_n \, .
\label{c8}
\ee
Thus the fundamental relation (\ref {c5}) not only
 yields the spectra of all $I_n$s, but
also connect them with the conserved quantities which appear 
in the framework of QISM as expansion coefficients of 
 $\ln {\hat A}(\lambda) $. 

Next, we use the fundamental relation (\ref {c5}) to
explicitly construct the first few quantum conserved quantities of DNLS model.
Commutation relations between the Jost solutions and basic field 
operators play a crucial role in this construction.
It is found that the first three quantum conserved quantities (number operator,
momentum operator and the Hamiltonian) are given by [6]
\bea
&& I_0 \equiv N = \int_{-\infty}^{+\infty}\psi^\dagger( x )\psi( x ) dx \, ,~~
 I_1 \equiv P = - i\int_{-\infty}^{+\infty}\psi^\dagger(x)
\partial_x\psi(x) \, dx \, , \nn \\
&& I_2 \equiv H = \int_{-\infty}^{+\infty}
\left\{ - \psi^\dagger(x) \d_{xx} \psi(x)
+  i\xi_q \, 
{\psi^\dagger}^2(x) \partial_x \psi^2(x) \right\}
dx \, ,  \nn ~~~~~~~~ (16a,b,c)
\eea
\addtocounter{equation}{1}
where
\be
\xi_q = \frac {\xi}{\sqrt {1- \hbar^2\xi^2}} \, . 
\label{c9}
\ee
It is worth noting that,
due to the  modification of coupling constant, the quantum
Hamiltonian (16c) can not be expressed as normal ordered version of the
corresponding classical Hamiltonian (\ref {a2}). 
Interestingly, the relation (\ref {c9})
between $\xi$ and $\xi_q$ is rather similar to the relation 
between rest mass and dynamical mass of a relativistic particle.
 Just as the dynamical mass
of a relativistic particle coincides with its rest mass in
 the nonrelativistic limit, 
 $\xi_q$  coincides with $\xi$ at $\hbar \rightarrow 0$ limit.
In the ultrarelativistic limit, 
 the dynamical mass of a particle tends towards infinity.
In a similar way, $\xi_q$ 
can take arbitrary large value at 
$ \v \xi \v  \rightarrow \frac{1}{\hbar}$ limit.
Consequently,
even though QYBE restricts the value of $\xi$ as 
 $ \v \xi \v  \leq \frac{1}{\hbar}$, there exists no such restriction
on the value of corresponding quantum coupling constant $\xi_q$ (\ref {c9}).
Thus the apparent limitation about the applicability of 
QISM in solving quantum DNLS Hamiltonian for the full range of its
coupling constant is resolved in a very nice way.

\section{Conclusion}
By applying a variant of quantum inverse scattering method, we derive all
possible commutation relations among the elements of the quantum monodromy
matrix associated with DNLS model. These commutation relations
enable us to construct the exact 
 eigenstates for all conserved quantities of quantum DNLS model through 
algebraic Bethe ansatz.  Next, we
propose the fundamental relation for DNLS model. Using this fundamental
relation, we are able to
explicitly construct the quantum Hamiltonian and few other
conserved quantities through basic field 
operators of this system. Surprisingly we find that,
 a new kind of coupling constant ($\xi_q$),  quite 
different from the classical one ($\xi$), appears in the quantum
Hamiltonian (16c) of the DNLS model. As a result, 
unlike the cases of most other integrable systems, this quantum
Hamiltonian (16c) can not be obtained as normal ordered version of the
corresponding classical Hamiltonian (\ref {a2}). 
Due to the presence of modified coupling constant
in the quantum Hamiltonian,
we are also able to consistently match various results of algebraic and
coordinate Bethe ansatz in the case of DNLS model [6]. The $S$-matrix
for two particle scattering and the distribution of 
single-particle momentum for 
quantum $N$-soliton states are two such examples where the results
of algebraic and coordinate Bethe ansatz match with each other.


\end{document}